\newif\ifarxiv
\newif\ifanonymous
\newif\ifpaper
\newif\ifTwoColumn
\papertrue

\arxivtrue

\ifarxiv
    \documentclass[acmsmall,screen,authorversion,nonacm]{acmart}
\else
    \documentclass[acmsmall,screen,review,anonymous]{acmart}
\fi

\makeatletter
\if@ACM@anonymous\anonymoustrue\else\anonymousfalse\fi
\ifnum\pdfstrcmp{\ACM@format}{sigconf}=0 \TwoColumntrue\else\TwoColumnfalse\fi
\makeatother

\usepackage[utf8]{inputenc}
\usepackage[T1]{fontenc}
\usepackage{microtype}
\usepackage{amsmath}
\usepackage[aboveskip=5pt, textfont=normalfont]{caption}
\usepackage{subcaption}
\usepackage{hyperref}

\usepackage{listings}
\usepackage[most]{tcolorbox}
\usepackage{fancyvrb}
\usepackage{changepage}
\usepackage{csvsimple}  
\usepackage{booktabs}   
\usepackage{array}      
\usepackage{multirow}
\usepackage{enumitem}
\usepackage{makecell}
\usepackage{tabularx}
\usepackage{array}
\usepackage{rotating}
\usepackage{cleveref}
\usepackage{siunitx}
\usepackage{algorithm}
\usepackage{algorithmicx}
\usepackage{algpseudocodex}

\usepackage{natbib}
\usepackage{balance}

\usepackage{tikz}

\usepackage{silence}
\ifanonymous
    \newcommand{\redacted}[1]{[redacted]}
\else
    \newcommand{\redacted}[1]{#1}
\fi

\newcommand{\punt}[1]{}

\newcommand{\dunder}[1]{\texttt{\_\_#1\_\_}}

\newcommand{\listingfont}{\ttfamily\scriptsize}

\definecolor{mygreen}{rgb}{0,0.6,0}
\definecolor{mygray}{rgb}{0.5,0.5,0.5}
\definecolor{mymauve}{rgb}{0.58,0,0.82}
\newtcblisting{pythonlisting}[1][]{
  beforeafter skip=0pt,
  coltitle=black,
  colbacktitle=gray!15,
  colback=gray!5,
  fonttitle={\small},
  boxrule=0pt,
  arc=0pt, 
  top=1pt,
  bottom=0pt,
  left=0pt,
  right=0pt,
  listing only,
  listing options={
    language=Python,
    numbers=none,
    morekeywords={as}
  },
  #1
}

\newtcblisting{errormessage}[1][]{
  before skip=0pt,
  after skip=0pt,
  coltitle=black,
  colback=yellow!15,
  colframe=black,
  boxrule=0pt,
  arc=0pt, 
  top=0pt,
  bottom=0pt,
  left=0pt,
  right=0pt,
  listing only,
  listing options={breaklines=true},
  #1
}

\newtcolorbox{groupbox}[1][]{
  beforeafter skip=0pt,
  coltitle=black,
  colbacktitle=gray!15,
  colback=blue!5,
  boxrule=0pt,
  arc=0pt, 
  top=1pt,
  bottom=0pt,
  left=0pt,
  right=0pt,
  #1
}

\newcommand{\concat}{\mathbin{\oplus}}
\newcommand{\ForEach}[1]{\For{\textbf{each} #1}}
\newcommand{\InCall}[2]{\textsc{#1}(#2)}
\newcommand{\In}[0]{\textbf{in }}

\newtcolorbox{conclusion}{%
    colback=green!5!white,
    colframe=green!75!black,
    arc=1mm,
    top=1mm,
    bottom=1mm,
    left=1mm,
    right=1mm
}

\usetikzlibrary{decorations.pathmorphing}

\newcommand{\pn}[2]{#2}

\newcommand{\pct}[1]{\SI{#1}{\%}}
\newcommand{\PCT}[1]{\textbf{\pct{#1}}}

\newcommand{\righttyperurl}{\redacted{\url{https://github.com/RightTyper/RightTyper}}}
\ifanonymous
    \newcommand{\replicationurl}{\url{https://anonymous.4open.science/r/righttyper-eval-0F82}}
\else
    \newcommand{\replicationurl}{\redacted{\url{https://github.com/plasma-umass/righttyper-eval}}}
\fi

\newcommand{\parheading}[1]{\subsubsection*{#1}}
\newcommand{\partopic}[1]{\subsubsection*{#1}}

\newcommand{\righttyper}{\textsc{RightTyper}}
\newcommand{\Thispaper}{This paper}

\title{Getting Python Types Right with \textsc{\righttyper{}}}


\author{Juan Altmayer Pizzorno}
\orcid{0000-0002-1891-2919}
\affiliation{%
    \institution{University of Massachusetts Amherst}
    \city{Amherst}
    \state{MA}
    \country{USA}
}
\email{jpizzorno@cs.umass.edu}

\author{Emery D. Berger\textsuperscript{\textdagger}}
\orcid{0000-0002-3222-3271}
\affiliation{
    \institution{University of Massachusetts Amherst / \\ Amazon Web Services}
    \city{Amherst}
    \state{MA}
    \country{USA}
}
\email{emery@cs.umass.edu}

\begin{document}


\begin{abstract}
    
Python type annotations enable static type checking, but most code remains untyped because manual annotation is time-consuming and tedious.
Past approaches to automatic type inference fall short: static methods struggle with dynamic features and infer overly broad types; AI-based methods are unsound and miss rare types; and dynamic methods impose extreme overheads (up to \pn{derived.mt_max_overhead_approx}{270}$\times$), lack important language support such as inferring variable types, or produce annotations that cause runtime errors.

\Thispaper{} presents \righttyper{}, a novel hybrid approach for Python that produces accurate and precise type annotations grounded in actual program behavior.
\righttyper{} grounds inference in types observed during actual program execution and combines these observations with static analysis and name resolution to produce substantially higher-quality type annotations than prior approaches.
Through principled, statistically guided adaptive sampling, \righttyper{} balances runtime overhead with the need to observe sufficient execution behavior to infer high-quality type annotations.
We evaluate \righttyper{} against static, dynamic, and AI-based systems on both synthetic benchmarks and real-world code, and find that it consistently achieves higher semantic similarity to ground-truth and developer-written annotations, respectively, while incurring only approximately \pn{benchmarks.overhead.RightTyper.geomean_overhead_pct}{27}\% runtime overhead.
\end{abstract}

\maketitle

\ifanonymous\else
    \begingroup\renewcommand\thefootnote{\textdagger}
        \footnotetext{Work done at Amazon Web Services.}
    \endgroup
\fi
\pagestyle{plain}

\section{Introduction}
\ifpaper
Python is now firmly established as one of the most popular programming languages.
Despite being dynamically typed, Python has supported static type annotations since version 3.5, released in 2015~\cite{pep484}.
\else
Even though it is dynamically typed, Python has supported static type annotations since version 3.5, released in 2015~\cite{pep484}.
\fi
While the Python runtime does not enforce these type annotations, separate tools such as \texttt{mypy} and \texttt{pyright}~\cite{mypy, pyright} use them to perform static type checking, catching bugs before deployment that would otherwise manifest as runtime errors.
Beyond static verification, type annotations improve code readability, enable better IDE support (code completion and refactoring), and support runtime validation frameworks (e.g., \texttt{pydantic} and \texttt{beartype}).
Collectively, these benefits can significantly reduce debugging time and improve software reliability.

Despite these many uses and a decade of language support, Python codebases remain largely unannotated.
A recent study of almost ten thousand popular Python project repositories found that only 7\% of them use type annotations at all.
Among those with annotations, only about 8\% of function arguments and return types are actually annotated~\cite{10.1145/3540250.3549114}.
This gap highlights the practical challenges developers face in manually adding annotations, especially to large codebases.
To address this, \emph{type inference} tools aim to automatically infer and insert type annotations, reducing manual effort and making static typing more accessible.

Past approaches to Python type inference fall into three broad categories: \emph{static}, \emph{AI-based}, and \emph{dynamic}.
Static approaches reason about code without executing it, using information such as program structure, operations, and control flow to infer types~\cite{static-program-analysis}.
However, they are limited by dynamic language features, either supporting only subsets of the language or leaving portions of the code unannotated~\cite{10.1145/1806596.1806598, 10.1007/978-3-642-03237-0_17}.
They also tend to be imprecise, as they conservatively over-approximate program behavior~\cite{ruby-dynamic}.
For example, if a function could theoretically operate on either an integer or a string, a static tool may reflect both in its inference, even if only one is ever actually used---diminishing the effectiveness of static checkers for catching bugs.

AI-based approaches operate on source code to estimate the likely types of program elements, most commonly by applying machine learning models.
These methods are better equipped than static approaches to handle the complexities of dynamic languages, but they sacrifice soundness: predicted types may not reflect actual program behavior, undermining the effectiveness of type checking through false positives or false negatives.
Some hybrid systems attempt to address this unsoundness by using static analysis to validate model predictions~\cite{typewriter, hityper}.
However, this validation can only discard incorrect type predictions, potentially leaving more of the code unannotated.
Moreover, these methods struggle with rare types, and many support only a limited type vocabulary, making them ill-equipped to handle user-defined or evolving types~\cite{type4py}.

In contrast, dynamic approaches observe types at runtime by executing the program and recording the actual values passed to and returned from functions~\cite{typewriter}.
This strategy trades static completeness for runtime fidelity: it requires code execution, but yields more precise and realistic types that reflect actual program behavior.
Unfortunately, existing dynamic type inference tools for Python---MonkeyType and PyAnnotate---fall short in key ways~\cite{monkeytype, pyannotate}.
Although both systems can employ sampling of function calls to reduce overhead (disabled by default in MonkeyType), their designs require that some instrumentation remains continuously active, contributing to runtime overhead.
MonkeyType further exhaustively scans all observed containers to infer element types and persists the collected information in a SQLite database, incurring substantial runtime and storage costs~\cite{sqlite}.
As a result, execution can slow down by up to \pn{derived.mt_max_overhead_approx}{270}$\times$, and logging may consume gigabytes of disk space per second.
PyAnnotate\footnote{Apparently no longer maintained; last updated in March 2021.} samples function calls and container elements, but its deterministic, ad hoc strategy can bias inferred annotations.
Moreover, PyAnnotate relies on Python's runtime type names, which can lead to annotations that cause runtime errors.
Both MonkeyType and PyAnnotate also lack support for inferring variable types.

\ifpaper
\subsection{Contributions}
\else
\parheading{Contributions}
\fi
\Thispaper{} proposes \righttyper{}, a novel hybrid type inference approach for Python.
\righttyper{} sidesteps pitfalls of static and AI-based methods by sampling types observed during actual program execution.
It employs adaptive sampling strategies to minimize runtime overhead while maintaining inference accuracy.
\righttyper{} monitors function calls using a Poisson process, in which observations occur only during brief, randomly timed capture windows, allowing most instrumentation to be disabled outside these windows.
For container types, \righttyper{} fully scans small containers, but applies Good–Turing estimation as a stopping criterion for larger ones, an estimator developed by Turing to estimate unseen events~\cite{good-turing, vanderWalt2021}.
Additionally, \righttyper{} tracks previously seen containers to avoid re-typing them, while continuing to sample to detect changes.
Together, these strategies achieve low overhead (overall $\sim$\pn{benchmarks.overhead.RightTyper.geomean_overhead_pct}{27}\%) while using statistically principled sampling to obtain representative coverage.

By combining observed types with information from static analysis, type aggregation, and name resolution, \righttyper{} emits annotations that substantially improve upon those produced by prior approaches.
\Thispaper{} makes the following contributions:

\begin{itemize}[topsep=5pt]
    \item Introduces \righttyper{}, a novel dynamic--static hybrid approach to type inference for Python (\S\ref{approach});
    \item Presents an open-source prototype implementation of \righttyper{} (\S\ref{implementation});
    \item Provides an enhanced implementation of TypeSim---a semantic type similarity metric from prior work~\cite{typybench}---addressing limitations in the original and released publicly (\S\ref{typesim});
    \item Demonstrates, through extensive empirical evaluation, that \righttyper{} produces substantially higher-quality type annotations than prior approaches while incurring significantly lower runtime overhead (\S\ref{eval-typeevalpy}, \S\ref{eval-real-world}, \S\ref{eval-performance});
    \item Shows that \righttyper{}’s adaptive sampling strategies achieve high type recall while substantially reducing sampling effort (\S\ref{eval-sampling}, \S\ref{eval-good-turing}).
\end{itemize}

\section{Approach}\label{approach}

\begin{figure*}
    \centering
    \includegraphics[width=\textwidth]{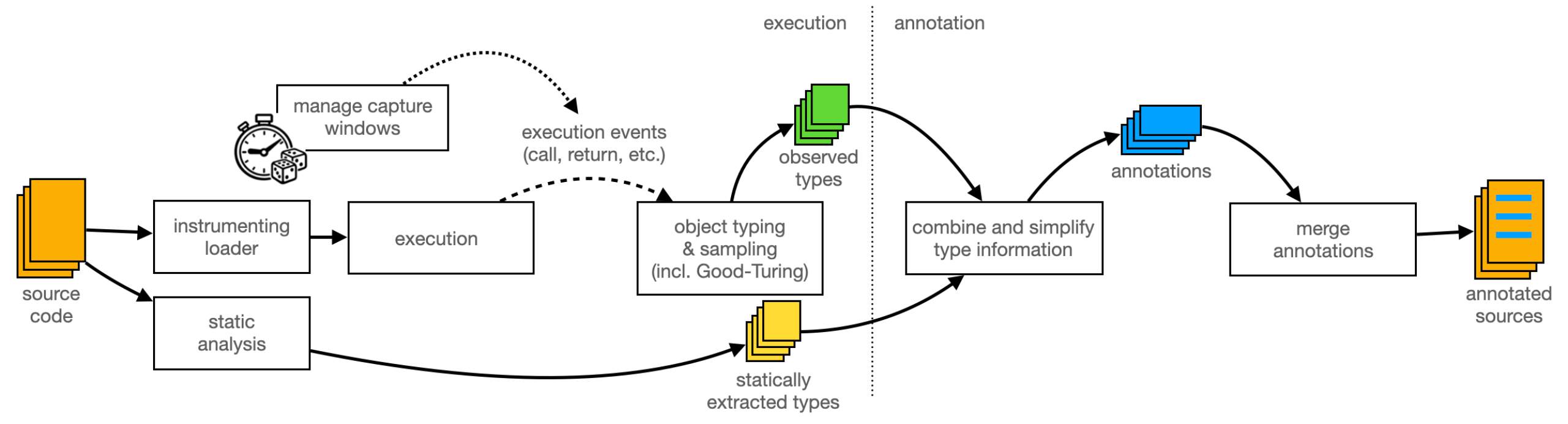}
    \caption{\textbf{\righttyper{} Overview:}
        \textnormal{
        \righttyper{} executes the target program under instrumentation,
        dynamically managing it to create Poisson-distributed observation windows (\S\ref{approach-instrumentation}).
        As the instrumentation delivers execution events, it observes and collects types used at runtime (\S\ref{object-typing}, \S\ref{container-sampling}).
        After execution, \righttyper{} combines these runtime observations with types extracted via static analysis to generate type annotations (\S\ref{typing-functions}--\S\ref{annotating-code}).
        }
    }
    \label{fig:overview}
\end{figure*}

\righttyper{} executes the target program using a combination of custom instrumentation and Python’s standard monitoring mechanisms.
During execution, it inspects runtime objects---such as function parameters, return values, and variables---to observe their types, and samples the contents of container objects (e.g., lists and dictionaries) to infer their complete types.
To manage overhead, \righttyper{} controls instrumentation using a Poisson process, enabling and disabling monitoring to create brief, randomly timed capture windows.
After execution, \righttyper{} combines these runtime observations with types extracted via static analysis, identifies recurring type patterns, and simplifies them to produce high-quality annotations.
Figure~\ref{fig:overview} provides an overview.

\subsection{Instrumentation and Event Sampling}\label{approach-instrumentation}
Most of \righttyper{}'s instrumentation utilizes \verb|sys.monitoring|, a low-overhead, fine-grained mechanism introduced in Python~3.12 for tracking various events during execution~\cite{pep669}\ifpaper\else, whose adoption \slipcover{} helped motivate (Chapter~\ref{chapter-slipcover})\fi.
This feature allows selective activation and deactivation of monitoring for specific portions of user code.
When \verb|sys.monitoring| does not provide the required events, \righttyper{} inserts additional instrumentation (\S\ref{implementation}).

To minimize runtime overhead while obtaining representative type observations, \righttyper{} governs execution event monitoring using a Poisson process.
Each function---or, more precisely, each code object, most of which correspond to functions---undergoes an initial warmup phase during which the first $k$ invocations (with default $k{=}5$) are captured unconditionally.
This warmup helps ensure sufficient observations for basic type inference, even for infrequently executed functions and very short executions.

After warmup, monitoring transitions to Poisson-timed sampling: a timer fires at exponentially distributed intervals (with configurable rate $\lambda$, default $\lambda{=}2\ \mathrm{Hz}$), enabling observation only during brief capture windows.
This approach yields statistically principled coverage while maintaining low average overhead, as the majority of function calls execute without instrumentation.
Together, the warmup phase and Poisson sampling ensure that \righttyper{} observes all executed functions and captures a representative sample of their invocations over time.
\righttyper{} combines the types of a function’s arguments, variables, and return values—observed separately across multiple execution events—into a \emph{call trace}, which it later uses to generate annotations.
By capturing all types involved in a call, call traces enable \righttyper{} to identify recurring type patterns and produce more precise annotations (see \S\ref{typing-functions}).

\subsection{Typing Objects}\label{object-typing}
As it processes monitoring events, \righttyper{} examines the objects passed into and returned from each function, as well as any variables, to infer their types.

\begin{figure}
\begin{pythonlisting}
    def visit_simple_stmt(self, node: Node) -> Iterator[Line]:
        ...
        prev_type: Optional[int] = None
        ...
\end{pythonlisting}
    \caption{\textbf{Capturing variable initializations:} 
        \textnormal{
            Although \righttyper{} primarily captures variable types at function exit, its loader also records constant-valued initializations.
            This technique enables correct typing of optionally typed variables, such as \texttt{prev\_type} in this excerpt from the \texttt{black} source code.
        }
        \label{fig:optional-variable}
    }
\end{figure}

\parheading{Variables}
\righttyper{} captures variable types, including object and class attributes, by inspecting their values when a function returns or unwinds.
This design reflects a performance trade-off: in Python, variables may change type on any assignment, making continuous tracking expensive.
Capturing values at function exit instead provides a practical balance between accuracy and overhead.
Static analysis performed by \righttyper{}'s loader supports this process by identifying the variables and attributes that should be tracked so they can be queried efficiently at function exit.
The loader also records any constants assigned to a variable when it is first introduced, which is particularly useful for correctly typing variables with optional values, as illustrated in Figure~\ref{fig:optional-variable}.

\parheading{Type Names and Import Paths}
Because most types---aside from a few built-ins---must be imported for use in annotations, \righttyper{} must determine the module and name under which each type is available; together, we refer to these here as the \emph{import path}.
Because Python’s introspection does not always provide usable import paths---especially for types implemented in C---\righttyper{} constructs a mapping from type objects to import paths, applying heuristics to identify a canonical import for each type.
This approach yields more natural and consistent annotations across the codebase (\S\ref{implementation}).
For commonly used built-in objects that lack importable type definitions, such as iterators, \righttyper{} falls back on structural subtyping, using protocols such as \verb|Iterator|~\cite{pep484}.

\subsection{Typing Generics}\label{container-sampling}\label{typing-generics}
When a type is generic, \righttyper{} also attempts to infer its type parameters.
While some generic instances expose their parameters directly, many others---such as standard container types---do not, and often contain heterogeneous elements; exhaustively enumerating their contents becomes very expensive for large instances.
\righttyper{} addresses this challenge through adaptive sampling, combining Good–Turing estimation with novelty detection to efficiently infer container element types.

\parheading{Containers}
\righttyper{} tracks containers and, upon first encountering a container or detecting a change, inspects its contents, distinguishing between small and large containers.
\righttyper{} fully scans small containers (of size $\le 32$ by default), while it analyzes large containers using probabilistic sampling with a Good–Turing stopping criterion.
Originally developed during World War II by Alan Turing and later formalized by I.~J.~Good, the Good–Turing estimator estimates the probability of unseen events~\cite{good-turing, vanderWalt2021}.

For large containers, after collecting a minimum number of samples (default 24), \righttyper{} computes the singleton ratio $\hat{r}$---the proportion of observed types that appear exactly once:
$$\hat{r} = \frac{\sum_{t \in C} \mathbf{1}[C(t) = 1]}{n}$$

\noindent
where $C$ is a counter mapping each observed type $t$ to its count $C(t)$, $n$ is the number of samples, and $\mathbf{1}[\cdot]$ is the indicator function.
When this ratio falls below a threshold $\tau$ (default $\tau = 0.05$) for all parameters (e.g., dictionary key and value types are counted separately), sampling terminates.
A hard cap (default 128 samples) further prevents pathological cases from dominating execution time.
\righttyper{} aggregates all types observed across inspections into a cumulative history, from which it generates annotations.

To handle containers that evolve over time, \righttyper{} tracks container size and employs probabilistic spot-checking.
When a previously-sampled container is revisited, a re-scan (or re-sampling) occurs if the container's size has changed, or a random spot-check discovers a novel type not present in the cumulative history.

Tracking containers is not only important for performance, but also for correctness: mutable containers are typically invariant with respect to their type parameters.
For example, if \righttyper{} observes a function taking a \verb|list[int]| and later a \verb|list[str]|, these observations should lead to a \verb@list[int|str]@ type only if they correspond to the same container instance; otherwise, they reflect containers with different element types and \righttyper{} should not merge them, since a \verb|list[int]| does not permit adding \verb|str| elements.

\parheading{Iterators}
Iterators pose special challenges for typing because their interface does not allow inspection of return values without destructively updating their state.
For iterators implemented in Python, whose iteration method calls are visible through \verb|sys.monitoring|, \righttyper{} begins with a partial type and updates it to a complete one once it observes an iteration.
Built-in iterators, however, do not trigger monitoring events and require special handling (\S\ref{implementation}).

\parheading{Numerical Arrays}
Numerical array shape mismatches are a common source of errors in machine learning programs~\cite{gradualtensorshapechecking}.
While \texttt{numpy} does not support array shape annotations at the time of writing, \righttyper{} can optionally include shape information using the format that \texttt{jaxtyping} supports, a package that enables runtime shape checking.
For instance, \righttyper{} would annotate a $2 \times 3$ \texttt{numpy} array as \texttt{Float64[ndarray, "2 3"]}.
When emitting such annotations for arrays, \righttyper{} also attempts to identify patterns in the shapes observed across traces, replacing them with variables.
Figure~\ref{fig:fig-numerical-array-shapes} shows an example.
\begin{figure}[t]
\begin{pythonlisting}
from jaxtyping import Float64
from numpy import ndarray as arr

def f(x: int, v: Float64[arr, "10 20"]) -> Float64[arr, "20"]:
    ...

def f(x: int, v: Float64[arr, "10 10"]) -> Float64[arr, "10"]:
    ...
\end{pythonlisting}
\begin{pythonlisting}[before skip=5pt]
def f(x: int, v: Float64[arr, "10 D1"]) -> Float64[arr, "D1"]:
    ...
\end{pythonlisting}
    \caption{\textbf{Annotation from shape pattern:}
        \textnormal{
            As it does for type patterns, when generating \texttt{jaxtyping}-style annotations, \righttyper{} identifies recurring patterns in the observed array shapes across traces (top section) and replaces them with variables (bottom section), in a manner inspired by Hindley–Milner–style type generalization.
            To the best of our knowledge, no other existing typing tools support inferring and annotating array dimensions in this way.
        }
        \label{fig:fig-numerical-array-shapes}
    }
\end{figure}

\subsection{Typing Functions}\label{typing-functions}
\begin{figure}
\begin{pythonlisting}
def add(a, b):
    return a + b

...
add(10.0, 20.0)
add("foo", "bar")
\end{pythonlisting}
\begin{pythonlisting}[before skip=5pt]
def add(a: float|str, b: float|str) -> float|str:
    return a + b
\end{pythonlisting}
\begin{errormessage}[before skip=0pt]
error: Unsupported operand types for + ("float" and "str")  [operator]
\end{errormessage}
    \caption{\textbf{Na\"{\i}ve typing with unions:}
        \textnormal{
            This example illustrates a function with interdependent argument and return types.
            The union of observed types (bottom), emitted by MonkeyType, is overly permissive and incorrectly allows mixed-type inputs that result in a type error.
        }
        \label{fig:a-plus-b}
    }
\end{figure}

After the target program finishes executing, \righttyper{} uses the collected call traces to generate type annotations for each function.
When the type of an argument or return value varies across traces, the simplest annotation is a \emph{union} of the observed types.
For instance, if a function receives a number argument in one sample and a string in another, it may be annotated as \texttt{f(a: float|str)}.
While unions offer a flexible way to describe multiple observed types, they can also be overly permissive, potentially leading to errors or false negatives during static analysis.
In particular, they may fail to capture essential relationships between arguments.
For example, Figure~\ref{fig:a-plus-b} illustrates the function \texttt{add}, which returns the sum of two numbers or the concatenation of two strings.
Na\"{\i}vely annotating both arguments as \texttt{float|str} permits invalid combinations -- such as one argument being a float and the other a string -- that would result in either a static type error or, if undetected, a runtime error.
In this example, because the arguments are used directly with the addition operator, \verb|mypy| can detect the error, issuing the message shown in Figure~\ref{fig:a-plus-b}.

For this reason, rather than immediately constructing unions, \righttyper{} first searches for patterns in the observed types.
When it detects consistent variability across argument or return value types, \righttyper{} introduces a \emph{type argument} to capture this shared variability, binding it to the concrete types encountered during execution.
The center portion of Figure~\ref{fig:a-plus-b-correct} illustrates the resulting annotation.
The search operates recursively, allowing it to detect patterns nested within type arguments.
For instance, a function that returns one of the elements of a list might be annotated as \texttt{f(l: list[T]) -> T}.
Algorithm~\ref{alg:generalize-call-traces} describes the process in more detail.
If the target Python version does not support type arguments, \righttyper{} instead defines a \emph{type variable} to achieve the same effect, albeit in a more verbose manner, as the bottom section of Figure~\ref{fig:a-plus-b-correct} shows.

\begin{algorithm}[t]
\caption{\textbf{Call trace generalization:}
    \textnormal{
        The algorithm examines the types observed at each \emph{position} (argument or return value) in a collection of call traces, generating a type for each corresponding position in the function signature.
        It transposes the call traces to group types by position and operates recursively, enabling it to identify patterns in the arguments of generic types.
        The symbol $\concat$ denotes concatenation.
    }
}
\label{alg:generalize-call-traces}
\begin{algorithmic}
\Function{Generalize}{$traces$}
    \Function{Rebuild}{$pos\_types$}
        \If{all $pos\_types$ are specializations of a generic $G$}
            \State $arg\_positions \gets $\Call{transpose}{\InCall{arguments}{$pos\_types$}}
            \State $arguments \gets [\ ]$
            \ForEach{$p$ \In $arg\_positions$}
                \State $arguments \gets arguments \ \concat$ \Call{Rebuild}{$p$}
            \EndFor
            \State \Return $G[arguments]$
        \ElsIf{$pos\_types$ occurs more than once}
            \State $T \gets$ assign or retrieve typevar for $pos\_types$
            \State \Return $T$
        \Else
            \State \Return \Call{Union}{$pos\_types$}
        \EndIf
    \EndFunction

    \State $signature \gets [\ ]$
    \ForEach{$p$ \In \Call{transpose}{$traces$}}
        \State $signature \gets signature \ \concat $ \Call{Rebuild}{$p$}
    \EndFor
    \State \Return $signature$
\EndFunction
\end{algorithmic}
\end{algorithm}

\begin{figure}
\begin{pythonlisting}[before skip=5pt]
def add[T: (float, str)](a: T, b: T) -> T:
    return a + b
\end{pythonlisting}
\begin{pythonlisting}[before skip=5pt]
from typing import TypeVar

rt_T1 = TypeVar("rt_T1", int, str)
def add(a: rt_T1, b: rt_T1) -> rt_T1:
    return a + b
\end{pythonlisting}
    \caption{\textbf{\righttyper{} recognizes type patterns:}
        \textnormal{
            recognizing that \texttt{add} (Figure~\ref{fig:a-plus-b}) is consistently called with either strings or numbers, \righttyper{} annotates the function using either a type argument (if supported by the target Python version; top portion) or a generated type variable (\texttt{rt\_T1}; bottom portion).
        }
        \label{fig:a-plus-b-correct}
    }
\end{figure}

\subsection{Typing Methods}\label{typing-methods}
Methods require special handling because their arguments and return values may reference the current instance, which could be a subtype of the defining class.
Consider Figure~\ref{fig:inherited-method}: when \verb|scale()| is invoked on a \verb|Line| object, annotating \verb|self: Line| would be incorrect, as the \verb|mypy| error in the figure demonstrates.
Instead, \righttyper{} annotates using the defining class, or \verb|Self| from the \verb|typing| module when targeting Python versions that support it~\cite{pep673}.

\begin{figure}
\begin{pythonlisting}
class Shape:
    ...
    def scale(self, factor):
        ...

class Line(Shape):
    ...

line = Line()
...
line = line.scale(1.10)
\end{pythonlisting}
\begin{pythonlisting}[before skip=5pt]
class Shape:
    ...
    def scale(self: Line, factor: float):
        ...
\end{pythonlisting}
\begin{errormessage}[before skip=0pt, listing options={breaklines=true}]
error: The erased type of self "line.Line" is not a supertype of its class "shape.Shape"  [misc]
\end{errormessage}
    \caption{\textbf{Inheritance complicates typing:}
        \textnormal{
            in the top section, class \texttt{Line} inherits the \texttt{scale()} method from its parent class, \texttt{Shape}.
            When \texttt{scale()} is invoked on a \texttt{Line} object, its \texttt{self} parameter refers to that object---an instance of \texttt{Line}.
            However, annotating \texttt{self: Line} (bottom section), as MonkeyType and PyAnnotate do, is incorrect, as demonstrated by the \texttt{mypy} error.
        }
        \label{fig:inherited-method}
    }
\end{figure}

\begin{figure}
\begin{pythonlisting}
class Shape:
    ...
    def draw(self: Self, medium: Medium) -> None:
        ...

class Line(Shape):
    ...
    def draw(self, medium):
        ...

line = Line()
...
line.draw(Screen())
\end{pythonlisting}
\begin{pythonlisting}[before skip=5pt]
class Line(Shape):
    ...
    def draw(self: Self, medium: Medium|Screen) -> None:
        ...
\end{pythonlisting}
    \caption{\textbf{Method overriding complicates typing:}
        \textnormal{
        In the top section, class \texttt{Line} overrides the \texttt{draw()} method inherited from its parent class, \texttt{Shape}.
        Although \texttt{Line.draw()} is only ever invoked with \texttt{Screen} as its \texttt{medium}, the overriding method must remain compatible with all argument types accepted by \texttt{Shape.draw()}.
        Accordingly, \righttyper{} annotates the parameter using the union of the observed type(s) and those accepted by the parent method (bottom section).
        MonkeyType and PyAnnotate only annotate using the types they observe, which leads to errors (not shown).
        }
        \label{fig:overriding-method}
    }
\end{figure}
Method arguments also require handling to avoid violating the Liskov Substitution Principle (LSP)~\cite{10.1145/62138.62141}.
Consider Figure~\ref{fig:overriding-method}, where \verb|Line| overrides \verb|draw| from \verb|Shape|.
Even though callers only invoke \verb|Line.draw| with \verb|Screen| as the \verb|medium| argument, annotating it as \verb|medium: Screen| would violate the LSP by excluding other valid media accepted by the base class.
To address this, \righttyper{} inspects parent classes to determine whether a method overrides one from a superclass, and if so, annotates the overriding method's arguments using the union of observed and inherited types.

To determine parent class annotations, \righttyper{} consults type stubs from \verb|typeshed| when necessary, since the vast majority of the standard library lacks inline type annotations~\cite{typeshed}.

\subsection{Simplifying Types}\label{simplifying-types}
If \righttyper{} cannot identify a consistent pattern across function argument or return value types (\S\ref{typing-functions}), or when typing variables, it constructs a union of the observed types at that position.

Before forming a union, \righttyper{} attempts to simplify the set of inferred types, counteracting the tendency of dynamic inference to produce large unions of concrete types.
It does so by merging types subsumed by others (e.g., merging \verb|list[int]| into \verb@list[int|str]@) and by replacing groups of concrete types with an appropriate common supertype.
To preserve semantic correctness, \righttyper{} verifies that the chosen supertype provides all methods and attributes shared by the replaced types.
For numeric types, \righttyper{} follows Python’s “numeric tower”; for example, observing both \verb|int| and \verb|float| results in the simplified type \verb|float|~\cite{pep484}.

\parheading{Simplifying for Human Use}
Despite the simplifications above, annotations can still be complex.
\righttyper{} offers options to trade precision for readability, such as collapsing repeated argument types (e.g., \verb|tuple[int, int, int]| to \verb|tuple[int, ...]|), limiting nesting depth, and capping union size (unions exceeding a configurable limit are collapsed to \verb|Any|).

\subsection{Handling Test-Borne Types}\label{handling-test-types}
When a test suite drives execution, it often invokes production code with test-specific objects such as mocks.
Such types should not appear in production annotations, as they do not reflect intended public APIs.
\righttyper{} automatically detects test modules and excludes types defined in them from inferred annotations.

\subsection{Annotating Source Code}\label{annotating-code}
The final step modifies source code to insert type annotations.
By default, \righttyper{} only annotates unannotated program elements, preserving existing annotations; an option allows overwriting them when desired.
When an inferred type requires a new import, \righttyper{} guards it using \verb|TYPE_CHECKING| and emits string annotations (Figure~\ref{fig:if-type-checking}).
Static type checkers evaluate code under the equivalent of \verb|TYPE_CHECKING=True| and resolve string annotations, so this approach avoids runtime import issues---particularly circular dependencies---while preserving full type checking.

\begin{figure}
\begin{pythonlisting}
from typing import TYPE_CHECKING, Never, Self
if TYPE_CHECKING:
    import astroid

    ...
    
    def visit_functiondef(self: Self, node: "astroid.FunctionDef") -> None:
        ...
\end{pythonlisting}
    \caption{\textbf{\righttyper{} avoids runtime import errors:} 
        \textnormal{
            \righttyper{} guards added imports with \texttt{TYPE\_CHECKING} and uses string annotations, avoiding runtime import issues while remaining compatible with static type checkers (excerpt from the \texttt{pylint} package).
        }
        \label{fig:if-type-checking}
    }
\end{figure}
\section{Implementation}\label{implementation}
\punt{
\begin{table}
    \small
    \setcellgapes{1pt}\makegapedcells
    \caption{\textbf{Differences by target Python version:} to optimize its use of the typing system, \righttyper{} annotates objects differently depending on the target Python version.
        \label{table:target-python-versions}
    }
    \begin{tabularx}{\columnwidth}{@{}r | X@{}}
        \hline
        \thead[r]{Version} & \thead[l]{Description} \\
        \hline
        3.9 & Earliest Python version supported by \righttyper{}\\
        3.10 & Uses e.g. \texttt{int|str|None} instead of \texttt{Optional[Union[int, str]]} \\
        3.11 & uses \texttt{typing.Self} and \texttt{typing.Never} \\
        3.12 & uses type arguments rather than type variables to express patterns (Fig.~\ref{fig:a-plus-b-correct})\\
        \hline
    \end{tabularx}
\end{table}
}

Our implementation of \righttyper{} comprises approximately 8{,}000 lines of Python code and is available as open-source software on GitHub\ifanonymous\else~\cite{righttyper-github}\fi.
This section describes key implementation details.

\parheading{Instrumentation}
\righttyper{} uses \verb|sys.monitoring| events (\verb|PY_START|, \verb|PY_RETURN|, \verb|PY_YIELD|, \verb|PY_UNWIND|, and \verb|CALL|) to capture function arguments, return values, yielded values, exception-driven unwinding, and call-site information.
Yielding temporarily suspends execution to produce a value before later resuming.
Generator functions and coroutines (including \verb|async| functions) both yield values, but coroutines may also receive values during execution; because \verb|sys.monitoring| does not expose these received values, \righttyper{} supplements monitoring with AST-based instrumentation, intercepting code loading to inject the necessary hooks before execution.

\parheading{Import Paths}
Python’s runtime type information does not always yield usable import paths, particularly for built-in and C-implemented types.
To address this, \righttyper{} constructs a mapping from type objects to import paths by scanning Python modules for type definitions.
When multiple import paths exist, it applies heuristics to select a canonical one, prioritizing public exports (via \dunder{all}), avoiding internal names, and preferring shorter paths within the same package.
This process produces more natural and consistent annotations across the codebase.

\parheading{Sampling}
Because most standard containers do not support efficient random sampling, \righttyper{} uses iterator-based sampling when necessary.
To support statistically correct $\mathcal{O}(1)$ sampling for dictionaries, it optionally replaces \texttt{dict} with a custom implementation that maintains elements in an auxiliary list via AST-based instrumentation.
This replacement is optional, as some programs rely on behavior specific to the standard \verb|dict|.

\parheading{Typing Iterators}\label{impl-builtin-iterators}
Built-in iterators do not emit \verb|sys.monitoring| events.
To handle these cases, \righttyper{} uses Python’s garbage collection module to recover the underlying iterable (e.g., a container) and infer element types, falling back to \verb|Iterator[Any]| when this is not possible.
It identifies such iterators using a maintained list of expressions known to produce them.

\parheading{Field Classes}
Classes defined with \verb|@dataclass|, \verb|attrs|, or \verb|NamedTuple| have auto-generated \verb|__init__| (or \verb|__new__|) methods whose source code is unavailable at instrumentation time.
\righttyper{} detects instantiation of such classes via the \verb|CALL| event, dynamically enables monitoring on their constructors, and maps observed argument types to field annotations.

\parheading{Python Version Support}
While it relies on \verb|sys.monitoring| and therefore requires Python~3.12 or later to execute, it can emit annotations targeting Python~3.9 and later.
Given the many extensions introduced since type annotations were first added to Python~\cite{typing-peps}, \righttyper{} adapts its annotations to the target Python version, for example using \texttt{int|str|None} instead of \texttt{Optional[Union[int, str]]} for Python 3.10 and later.

\parheading{Libraries}
The implementation uses Python’s standard \verb|ast| module for AST-based instrumentation and leverages the \verb|libcst| concrete syntax tree library for static analysis and source-to-source transformation~\cite{libcst}.
It also uses \verb|typeshed_client| to access the \verb|typeshed| type stub repository~\cite{typeshed-client, typeshed}, which is necessary because much of the standard library and many third-party packages lack inline type annotations.

\newif\ifusecharts
\usechartstrue  

\section{Evaluation}
Evaluating type inference is a challenging task.
One of the difficulties lies in establishing a reliable ground truth.
Some prior work derives ground truth from developer-written annotations, sometimes supplemented with types inferred by static analyzers such as \texttt{pytype} or Pyre~\cite{typilus, type4py}.
However, many projects contain type errors, as type checking is not yet a standard part of development workflows, and relying on developer-written annotations restricts evaluation to already-typed codebases~\cite{10.1145/3540250.3549114}.
Moreover, types inferred by static tools are only as reliable as the tools themselves: due to dynamic language features, these tools may conservatively over-approximate types or fail to infer them altogether (\S\ref{static-type-inference}).
Existing datasets also often omit support for code execution, as they target static and AI-based type inference methods~\cite{manytypes4py, typeevalpy, typybench}.

To address these challenges, we evaluate \righttyper{} both on TypeEvalPy---a benchmark with an explicit ground truth, which we extend to support code execution---and on well-maintained real-world codebases, where inferred annotations can be compared against developer-written ones~\cite{typeevalpy}.

Another difficulty lies in comparing inferred types.
Prior work has often relied on string comparison of annotations~\cite{typewriter, manytypes4py, typilus, typeevalpy}.
But a program element containing a \verb|list[int]|, for example, can be validly annotated in multiple ways.
In addition to trivial aliases like \verb|List[int]| and less precise forms like \verb|list|, it may also be appropriate to use protocol-based types such as \verb|Collection|, \verb|Sequence|, or \verb|Iterable|.
Because string comparison cannot recognize semantic relationships between types, as the nesting depth increases (e.g., \verb@dict[str, str|int|list[str|int]]@), string match scores drop to 0 as it is unable to recognize partial matches~\cite{typybench}.
An effective comparison metric must be sufficiently nuanced to recognize when variants are appropriate and also indicate partial matches.
Some prior work instead adopts a ``correctness modulo type checker'' approach to evaluation, measuring the quality of inferred types on whether they pass static type checking without errors~\cite{10.1145/3689783, typilus, 10.1145/3485488}.
However, doing so reinforces the often unwarranted assumption that the underlying code is correct.
As Figure~\ref{fig:quac-silences-mypy} illustrates, annotations can also effectively silence type checkers, inadvertently concealing bugs.
For this evaluation, we use a modified version of the TypeSim type similarity metric proposed in prior work, which enables comparison with both established ground truth and developer-written annotations (\S\ref{typesim}).
We do not use TypyBench's TypeCheck metric, which counts \verb|mypy| errors in annotated code.
Dynamic approaches only annotate executed code, leaving unexecuted functions unannotated; when annotated code calls unannotated code (or vice versa), \verb|mypy| reports type errors at these call sites even if the annotations themselves are correct.
TypeCheck thus penalizes dynamic tools for incomplete coverage rather than annotation quality.
\begin{figure}
\begin{pythonlisting}
def add(a, b):
    return a + b

add(10, 20)
add("foo", "bar")
\end{pythonlisting}
\begin{pythonlisting}[before skip=5pt]
from _typeshed import SupportsAdd
def add(a: SupportsAdd, b: SupportsAdd) -> SupportsAdd:
    return a + b
\end{pythonlisting}
\begin{pythonlisting}[before skip=5pt]
add(10, "bar")
\end{pythonlisting}
\begin{errormessage}
TypeError: unsupported operand type(s) for +: 'int' and 'str'
\end{errormessage}
    \caption{\textbf{QuAC silences \texttt{mypy}:}
        \textnormal{
        The code, previously shown in Figure~\ref{fig:a-plus-b}, contains a function that, when invoked with two numbers, returns their sum, but when invoked with two strings, returns their concatenation.
        QuAC infers the \texttt{SupportsAdd} protocol, which prevents \texttt{mypy} from detecting the type mismatch error (middle).
        Executing the program yields a Python runtime error (bottom).
        }
        \label{fig:quac-silences-mypy}
    }
\end{figure}

\medskip\noindent
Our evaluation investigates the following questions:
\begin{description}
    \item[\textbf{RQ1:}] Do type annotations generated by \righttyper{} improve upon those produced by prior approaches?
    (\S\ref{eval-typeevalpy}, \S\ref{eval-real-world})
    \item[\textbf{RQ2:}] How does \righttyper{}'s runtime overhead compare with that of other dynamic approaches?
    (\S\ref{eval-performance})
    \item[\textbf{RQ3:}] How effective is \righttyper{}'s sampling approach?
    (\S\ref{eval-sampling})
    \item[\textbf{RQ4:}] How effective is Good--Turing estimation as a stopping criterion for container sampling?
    (\S\ref{eval-good-turing})
\end{description}

\smallskip
\noindent
Before addressing these questions, we describe the TypeSim metric used in our evaluation.

\subsection{The TypeSim Metric and Its Implementation}\label{typesim}
We compare annotations using a modified version of the TypeSim metric proposed by TypyBench~\cite{typybench}.
TypeSim provides a continuous measure of similarity between types based on the operations they support.
It compares base types using the Jaccard index, defined as the ratio of shared attributes to the total number of attributes across both types:

$$\operatorname{Jaccard}(t, t') = \frac{|\operatorname{attrs}(t)\cap\operatorname{attrs}(t')|}{|\operatorname{attrs}(t)\cup\operatorname{attrs}(t')|}$$

TypeSim operates recursively to support generic types.
It handles unions by first computing an optimal matching between their elements and then comparing the matched elements.

Because the original implementation builds on \verb|mypy|, the types it compares reflect not only written annotations but also types inferred by mypy from context, conflating annotation accuracy with inference behavior. In addition, it compares only attributes declared directly in each class body rather than the full set of inherited attributes, which can yield a similarity of zero between a class and its parent.
To avoid these issues, we reimplement TypeSim using \verb|libcst| and operate directly on annotation syntax.

We further extend TypeSim to support commonly used typing constructs---such as \verb|Callable|, \verb|TypedDict|, \verb|Literal|, and type aliases---that the original implementation ignored and therefore excluded from evaluation.
Our version evaluates all of these constructs: callable types are compared structurally, literal types receive partial credit when they share a base type, and type aliases and type variables are expanded during extraction rather than rejected at comparison time.
We also normalize semantically equivalent constructs before comparison---for example, \verb|Generator[X, None, None]| is reduced to \verb|Iterator[X]|---so that superficial differences in spelling do not penalize otherwise correct predictions.

As in the original TypeSim, and following prior work, we exclude top-level \verb|Any| annotations from evaluation because they are uninformative~\cite{10.1145/3689783, typybench}.
For consistency, we also exclude top-level \verb|object| annotations and assign a score of zero to \verb|Any| and \verb|object| when they appear within compound types (e.g., \verb@dict[str, Any]@).
The original implementation could assign full credit in such cases through string matching, even when part of the type structure was uninformative.

\subsection{[RQ1] Type Inference on TypeEvalPy}\label{eval-typeevalpy}
We address RQ1 by first evaluating \righttyper{} on TypeEvalPy, a benchmark suite of 153 short programs designed to evaluate type inference across a range of Python language features~\cite{typeevalpy}.
To make this possible, we add explicit calls to ensure code execution and adjust its assertions to target only program elements for which Python permits annotations (excluding, for example, lambda parameters), resulting in a total of \pn{typeevalpy.total}{827} type assertions.

We then evaluate \righttyper{} on the modified suite.
In doing so, we identify a number of discrepancies, all of which we determined to be errors in the benchmark’s ground-truth files.
We corrected these errors and submitted the fixes to the benchmark authors, who accepted them.
Finally, we add support for running QuAC, MonkeyType, pytype, and Type4Py as baselines.
Given its apparent lack of maintenance, we do not include PyAnnotate in our evaluation.
We also exclude Pyre, as its \verb|infer| command is too limited to be useful: it cannot infer argument types from usage, return types from operations, or types for class methods.
We execute MonkeyType without enabling sampling, allowing it to record every execution and thus achieve its best possible results.
TypeEvalPy already supports prompting OpenAI models for type inference; we include results obtained by prompting the GPT-4o model using that implementation.
Despite our best efforts to correct them, bugs in TypeT5 prevent us from including it in this comparison.
RunTyper~\cite{10.1145/3771544} was also excluded as it requires a modified Python interpreter incompatible with standard installations.

\ifusecharts Figure~\ref{table:typeevalpy}(a) \else Table~\ref{table:typeevalpy} \fi presents the results in terms of TypeSim scores\ifusecharts\else, the proportion of ``exact'' matches achieving 100\% TypeSim, and coverage, measured as the fraction of ground-truth items for which inferences are generated\fi.
With an overall \pn{typeevalpy.tools.RightTyper.typesim}{99.8}\% TypeSim, \righttyper{} achieves a near-perfect score, with only two cases failing to yield an exact match.
In one case, the benchmark requests the type of an abstract method that does not execute.
In the other case, \righttyper{} correctly types a \verb|zip| iterator using the \verb|Iterator| protocol---necessary to specify element types, since \verb|zip| itself does not support them---but receives a lower score.
This lower score occurs because TypeSim penalizes protocol-based annotations: protocols expose only the minimal set of required operations and therefore appear less similar to richer concrete types under the metric.
GPT-4o, at \pn{typeevalpy.tools.GPT-4o.typesim}{95}\%, also performs well, only incorrectly inferring ``Unknown’’ or \verb|Any| in a few cases.
It is followed by pytype at \pn{typeevalpy.tools.pytype.typesim}{60.6}\%; it suffers from low coverage, consistent with prior work showing that its conservative static analysis leads to limited coverage~\cite{10.1145/3652153}.
MonkeyType and QuAC achieve \pn{typeevalpy.tools.MonkeyType.typesim}{32.9}\% and \pn{typeevalpy.tools.QuAC.typesim}{23}\%, respectively, in large part because they lack support for inferring variable types\ifusecharts{} (Figure~\ref{fig:breakdown})\fi.
QuAC operates by collecting sets of required attributes and using BM25---a term-frequency ranking function from information retrieval---to match these sets to candidate types~\cite{bm25}.
It is therefore plausible that the benchmark’s short programs---designed to facilitate manual ground-truth verification---provide limited context for this matching process, reducing QuAC’s effectiveness~\cite{10.1145/3689783}.
Similarly, Type4Py’s relatively low performance (\pn{typeevalpy.tools.Type4Py.typesim}{24.8}\%) may be due to the benchmark programs likely representing out-of-distribution inputs for its model.

\ifusecharts
\begin{figure*}[t]
    \centering
    \begin{minipage}[t]{0.48\textwidth}
        \centering
        \includegraphics[width=\linewidth]{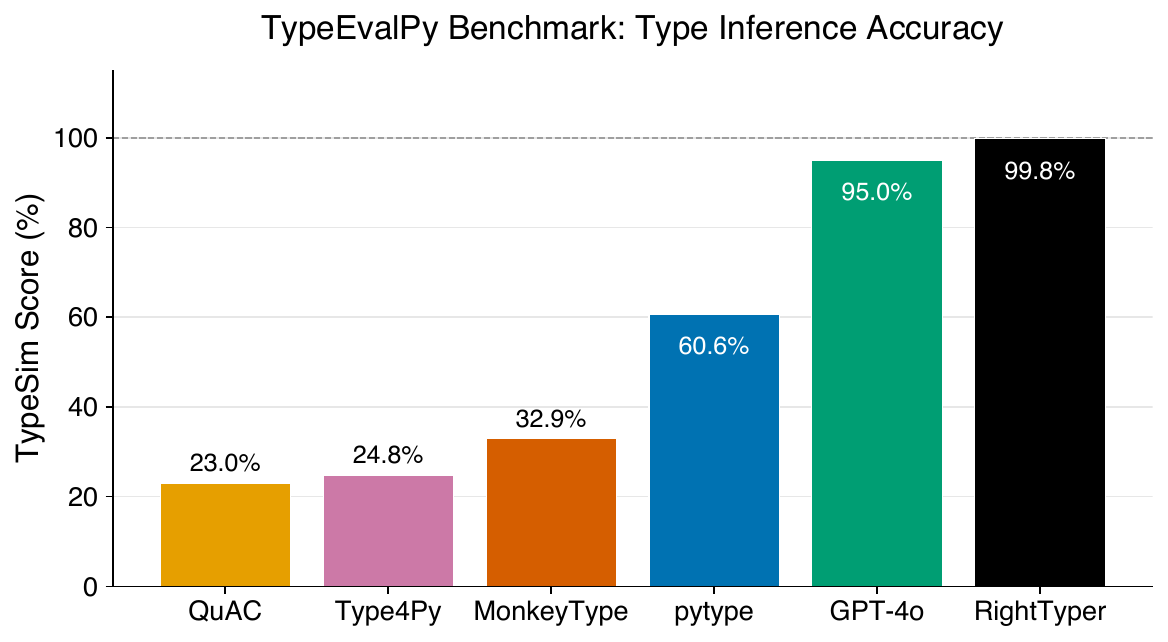}
        \subcaption{TypeEvalPy benchmark (\pn{typeevalpy.total}{827} annotations)}
        \label{fig:typeevalpy}
    \end{minipage}
    \hfill
    \begin{minipage}[t]{0.48\textwidth}
        \centering
        \includegraphics[width=\linewidth]{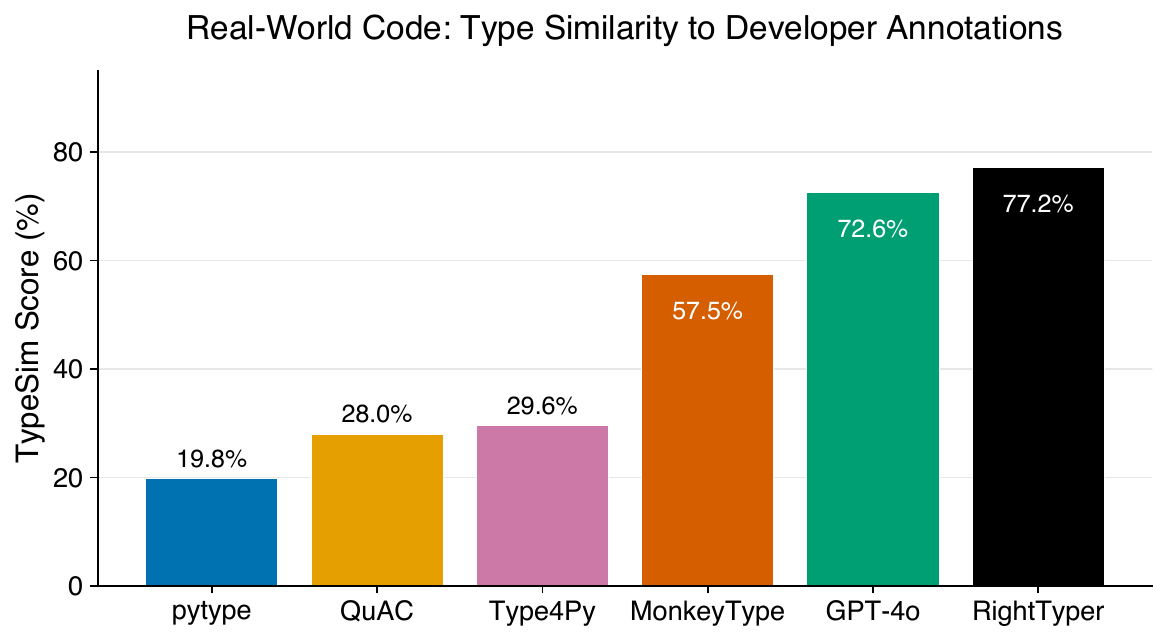}
        \subcaption{Real-world code (\pn{realworld.total}{10{,}964} annotations)}
        \label{fig:realworld}
    \end{minipage}
    \caption{
        \textbf{[RQ1] \righttyper{} achieves the highest type inference accuracy}:
        (a) On TypeEvalPy, \righttyper{} achieves \pn{typeevalpy.tools.RightTyper.typesim}{99.8}\% TypeSim, outperforming GPT-4o (\pn{typeevalpy.tools.GPT-4o.typesim}{95}\%).
        (b) On real-world code, \righttyper{} achieves \pn{realworld.tools.RightTyper.typesim}{77.2}\% TypeSim, outperforming GPT-4o (\pn{realworld.tools.GPT-4o.typesim}{72.6}\%) and MonkeyType (\pn{realworld.tools.MonkeyType.typesim}{57.5}\%).
    }
    \label{table:typeevalpy}
    \label{table:real-world}
\end{figure*}
\else
\begin{table*}
    \caption{
        \textbf{[RQ1] \righttyper{} excels on TypeEvalPy}: \righttyper{} achieves a near-perfect result, with only two of 827 types failing to yield an exact match.
        \label{table:typeevalpy}
    }
\begin{tabular}{c l@{\hspace{4em}} r r r r r r}
\toprule
& Metric & QuAC & Type4Py & MonkeyType & pytype & GPT-4o & RightTyper \\
\midrule
\multirow{3}{*}{\rotatebox{90}{\scriptsize overall}} & TypeSim & \pct{23.0} & \pct{24.8} & \pct{32.9} & \pct{60.6} & \pct{95.0} & \PCT{99.8} \\
 & Exact Match & \pct{22.9} & \pct{20.9} & \pct{32.8} & \pct{59.0} & \pct{94.8} & \PCT{99.8} \\
 & Coverage & \pct{25.6} & \pct{78.6} & \pct{32.9} & \pct{72.9} & \PCT{100.0} & \pct{99.9} \\
\midrule
\multirow{3}{*}{\rotatebox{90}{\scriptsize funcs.}} & TypeSim & \pct{51.2} & \pct{17.7} & \pct{73.0} & \pct{46.9} & \pct{95.8} & \PCT{99.7} \\
 & Exact Match & \pct{50.8} & \pct{14.5} & \pct{72.8} & \pct{46.5} & \pct{95.4} & \PCT{99.7} \\
 & Coverage & \pct{57.0} & \pct{56.7} & \pct{73.1} & \pct{63.2} & \PCT{100.0} & \pct{99.7} \\
\midrule
\multirow{3}{*}{\rotatebox{90}{\scriptsize vars.}} & TypeSim & \pct{0.0} & \pct{30.5} & \pct{0.0} & \pct{71.7} & \pct{94.3} & \PCT{99.8} \\
 & Exact Match & \pct{0.0} & \pct{26.2} & \pct{0.0} & \pct{69.2} & \pct{94.3} & \PCT{99.8} \\
 & Coverage & \pct{0.0} & \pct{96.5} & \pct{0.0} & \pct{80.9} & \PCT{100.0} & \PCT{100.0} \\
\bottomrule
\end{tabular}
\\[0.5em]
\small Results on 827 type annotations (372 functions, 455 variables).
\end{table*}

\fi

\subsection{[RQ1] Type Inference on Real-World Code}\label{eval-real-world}
\ifusecharts
\else
\begin{table*}[t]
\centering
    \caption{\textbf{[RQ1] \righttyper{}-inferred types are most similar to developers'}: Evaluated on real-world code, \righttyper{} achieves the highest TypeSim scores.
}
\label{table:real-world}
\begin{tabular}{c l@{\hspace{4em}} r r r r r r}
\toprule
& Metric & Type4Py & QuAC & MonkeyType & pytype & GPT-4o & RightTyper \\
\midrule
\multirow{3}{*}{\rotatebox{90}{\scriptsize overall}} & TypeSim & 29.6\% & 28.0\% & 57.5\% & 19.8\% & 72.6\% & \textbf{77.2\%} \\
 & Exact Matches & 22.6\% & 20.0\% & 50.1\% & 17.5\% & \textbf{65.1\%} & 64.2\% \\
 & Coverage & 78.1\% & 45.2\% & 66.3\% & 24.3\% & 81.4\% & \textbf{93.1\%} \\
\midrule
\multirow{3}{*}{\rotatebox{90}{\scriptsize functions}} & TypeSim & 30.9\% & 31.0\% & 63.7\% & 21.0\% & 78.5\% & \textbf{78.6\%} \\
 & Exact Matches & 23.6\% & 22.1\% & 55.5\% & 18.9\% & \textbf{70.6\%} & 65.9\% \\
 & Coverage & 82.3\% & 50.1\% & 73.5\% & 25.2\% & 87.7\% & \textbf{94.2\%} \\
\midrule
\multirow{3}{*}{\rotatebox{90}{\scriptsize variables}} & TypeSim & 17.4\% & 0.0\% & 0.0\% & 8.6\% & 17.6\% & \textbf{64.5\%} \\
 & Exact Matches & 13.5\% & 0.0\% & 0.0\% & 4.5\% & 14.5\% & \textbf{48.5\%} \\
 & Coverage & 38.8\% & 0.0\% & 0.0\% & 16.0\% & 23.6\% & \textbf{83.5\%} \\
\bottomrule
\end{tabular}
\\[0.5ex]
{\small Results on 10{,}964 annotations (9{,}894 functions, 1{,}070 variables); larger results are better.}
\end{table*}

\fi
Although the programs in TypeEvalPy provide broad coverage of Python language features, they are significantly simpler than, and differ markedly from, commonly deployed Python code.
We therefore next evaluate \righttyper{} on the real-world Python packages \texttt{black}, \texttt{flask}, \texttt{httpx}, \texttt{pylint}, and \texttt{rich}, which together contain over \pn{derived.realworld_total_approx_k}{10}{,}000 annotations.
We select these packages based on several criteria.
First, all five are well-annotated open-source Python projects, with annotations developed and refined by expert maintainers over many years of use with type checkers---making them ideal ground-truth sources.
Second, they are complex, production-quality codebases that exercise diverse Python features including inheritance hierarchies, generic types, protocols, decorators, context managers, and AST manipulation.
Third, they are widely used (\verb|pepy.tech| reports billions of downloads) and well-maintained, reducing the risk of ground-truth errors.
Finally, the packages span command-line tools, libraries, and a web framework, allowing us to drive them end to end with varied inputs, achieving broader coverage than test suites alone.
As ground truth, we use the projects’ original type annotations; that is, we evaluate the semantic similarity of inferred annotations to developer-written ones using TypeSim.

Prior to evaluating each system, we remove all type annotations from the source code, except for special constructs such as \verb|dataclasses| and \verb|NamedTuple| where removing annotations would alter program semantics; in these cases, we replace the annotations with a type alias placeholder, allowing us to distinguish them from tool-inferred types during evaluation.
We exclude annotations originally typed as \verb|Any|, since they carry no type information and cannot meaningfully distinguish tool quality.
We also remove any docstrings, as these can influence AI-based type inference; this ensures a level comparison with non-AI methods that cannot leverage natural language hints.

When evaluating \righttyper{} and MonkeyType, we execute each target package multiple times with varying options and inputs---keeping them identical across the two systems---and also run their test suites to further exercise production code, thereby collecting typing information across a range of execution scenarios.
\righttyper{} excludes test functions from tracing by default; we configure MonkeyType to do the same, so that only production code is typed.
As before (\S\ref{eval-typeevalpy}), we run MonkeyType without sampling, allowing it to record every execution and achieve its best possible results.
Rather than having MonkeyType annotate the source code directly, we configure it to generate stub files, as in-source annotation causes MonkeyType to emit import statements that introduce circular dependencies.
Applying MonkeyType and QuAC to these sources uncovered several bugs, including an assertion failure in QuAC and syntax errors in MonkeyType’s output; we corrected these issues.
To include LLM-based results, we reimplement querying GPT-4o using TypyBench’s ``single-file context'' prompt, as its replication package does not provide an implementation.

\ifusecharts Figure~\ref{table:typeevalpy}(b) \else Table~\ref{table:real-world} \fi presents the results\ifusecharts\else; as before, we report TypeSim scores, the proportion of ``exact'' matches achieving 100\% TypeSim, and coverage, measured as the fraction of originally annotated program elements for which types are inferred\fi.
With an overall TypeSim score of \pn{realworld.tools.RightTyper.typesim}{77.2}\%, \righttyper{} achieves the highest similarity to developer-annotated types.
It is followed by GPT-4o and MonkeyType, which at \pn{realworld.tools.GPT-4o.typesim}{72.6}\% and \pn{realworld.tools.MonkeyType.typesim}{57.5}\% also show good performance, and at a considerable distance by Type4Py at \pn{realworld.tools.Type4Py.typesim}{29.6}\%, QuAC at \pn{realworld.tools.QuAC.typesim}{28}\%, and pytype at \pn{realworld.tools.pytype.typesim}{19.8}\%.

\ifusecharts
\begin{figure*}[!t]
    \centering
\ifpaper
    \begin{minipage}[t]{0.48\textwidth}
        \centering
        \includegraphics[width=\linewidth]{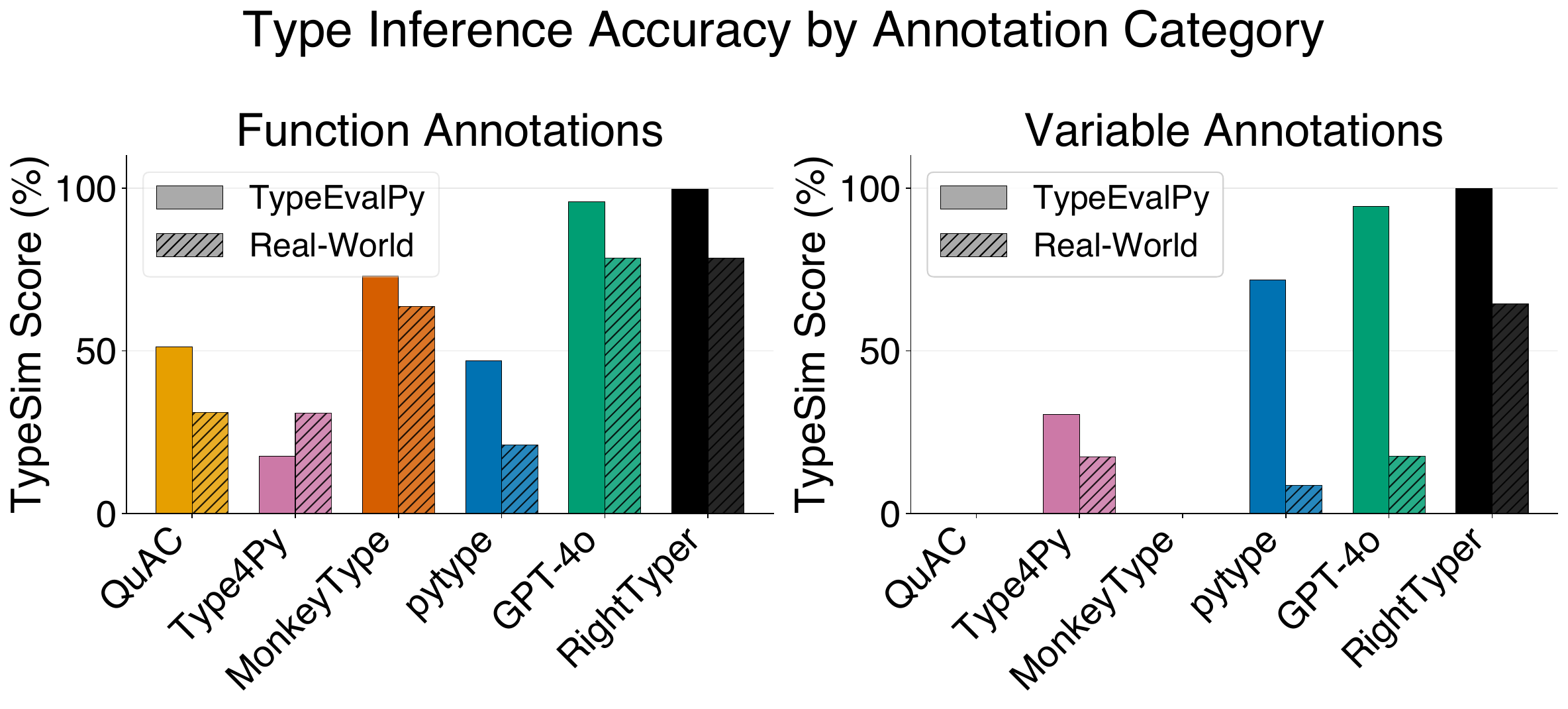}
        \subcaption{Functions vs.\ variables breakdown}
        \label{fig:breakdown}
    \end{minipage}
    \hfill
    \begin{minipage}[t]{0.48\textwidth}
        \centering
        \includegraphics[width=\linewidth]{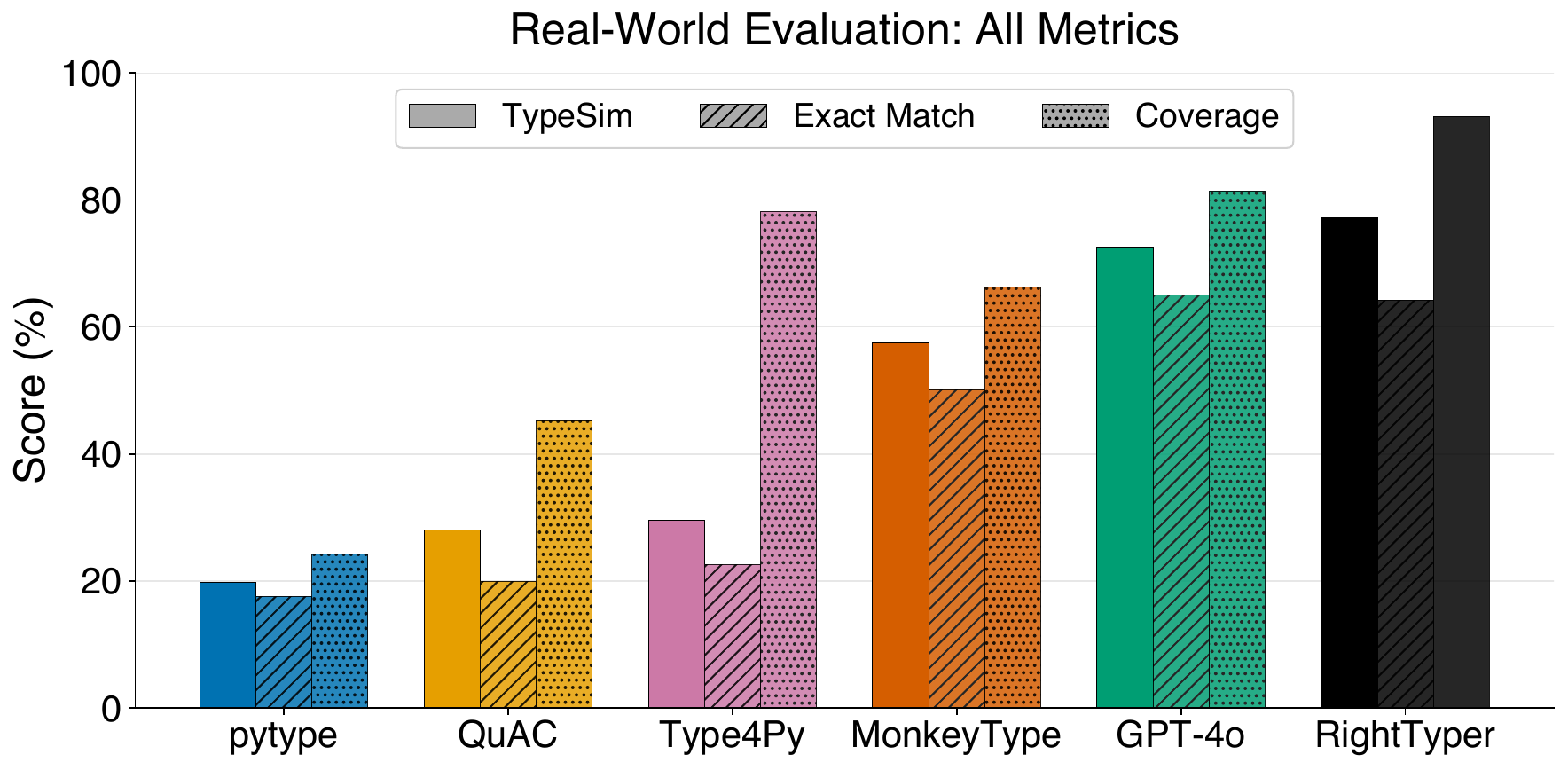}
        \subcaption{All metrics on real-world code}
        \label{fig:metrics}
    \end{minipage}
\else
    \includegraphics[width=0.85\textwidth]{figures/breakdown_chart.pdf}
    \subcaption{Functions vs.\ variables breakdown}
    \label{fig:breakdown}
    \vspace{1em}
    \includegraphics[width=0.7\textwidth]{figures/metrics_comparison.pdf}
    \subcaption{All metrics on real-world code}
    \label{fig:metrics}
\fi
    \caption{
        \textbf{[RQ1] Detailed results}:
        (a) TypeSim scores by annotation category show \righttyper{} excels on both functions and variables, while QuAC and MonkeyType cannot infer variable types.
        (b) On real-world code, \righttyper{} achieves the highest TypeSim (\pn{realworld.tools.RightTyper.typesim}{77.2}\%) and Coverage (\pn{realworld.tools.RightTyper.coverage}{93.1}\%).
    }
    \label{fig:rq1-details}
\end{figure*}
\fi

Examining \texttt{pytype}'s annotations, we find that they often omit type arguments and exhibit low coverage, consistent with the results in \S\ref{eval-typeevalpy}.
QuAC likewise suffers from low coverage, failing to infer types for nearly half of the function annotations---a limitation exacerbated by its lack of support for variable typing.
Its annotations also often confuse nodes with similar attribute sets, a consequence of its BM25-based attribute-to-type mapping.

Type4Py’s annotations reveal frequent prediction errors, including a bias toward common types such as \verb|str|, confusion between similarly named types from different libraries, omission of \verb|None| for optional parameters, and difficulty handling project-specific types.
These issues explain its low TypeSim score despite relatively high coverage.

Several factors affect MonkeyType's score: it often infers overly concrete types where the original annotations use protocols (e.g., \verb|List[str]| instead of \verb|Sequence[str]|), misses some code paths, and lacks support for variable typing.
GPT-4o performs relatively well on function annotations but poorly on variable annotations, which significantly lowers its overall score; it also frequently emits \verb|None| for optional types or falls back to \verb|Any| where more specific types were originally annotated.

\righttyper{} exhibits some similar limitations, including occasional overly concrete typing and missed coverage, the latter being expected given that it executes the same scenarios as MonkeyType.
However, \righttyper{} achieves higher scores overall due to more complete language support---such as inferring variable types, handling keyword and variadic arguments, inferring \verb|Callable| argument types, and handling nested functions---all of which MonkeyType lacks---as well as its ability to better simplify inferred types, yielding cleaner annotations that more closely match the originals\ifusecharts{} (Figure~\ref{fig:metrics})\fi.
The lower scores on variable annotations stem from variables captured only at function exit (missing intermediate types), unexecuted branches, and developers annotating with broader types than observed at runtime.
Finally, in some cases \righttyper{} correctly emits annotations with type parameters (cf.~\S\ref{typing-functions}), but TypeSim yields a zero score because it does not capture these forms of semantic equivalence.

\punt{Two additional factors affect scores.
First, platform-specific code that is unreachable in the test environment cannot be traced by dynamic tools: for example, \texttt{rich}'s Win32 console module raises \texttt{ImportError} on non-Windows platforms, leaving its annotations invisible to every dynamic tool.
Second, no tool successfully infers \texttt{Literal} types---\texttt{Literal} annotations across all five packages are inferred as their base types (\texttt{str}, \texttt{int}, \texttt{bool}), and the majority score below 0.5 in TypeSim.
}

Given \righttyper{}'s strong results on both TypeEvalPy and real-world code, we conclude that it improves annotation quality relative to prior approaches.

\begin{conclusion}
\textbf{[RQ1] Summary:} With near-perfect inference on TypeEvalPy and the highest TypeSim on real-world code (\pn{realworld.tools.RightTyper.typesim}{77.2}\% vs.\ \pn{realworld.tools.GPT-4o.typesim}{72.6}\% for GPT-4o and \pn{realworld.tools.MonkeyType.typesim}{57.5}\% for MonkeyType), \righttyper{} produces annotations of higher quality than prior approaches---enabled by grounding types in actual runtime behavior and supporting a wider range of Python language features.
\end{conclusion}

\subsection{[RQ2] Runtime Performance}\label{eval-performance}
\begin{figure*}
    \centering
    \includegraphics[width=\linewidth]{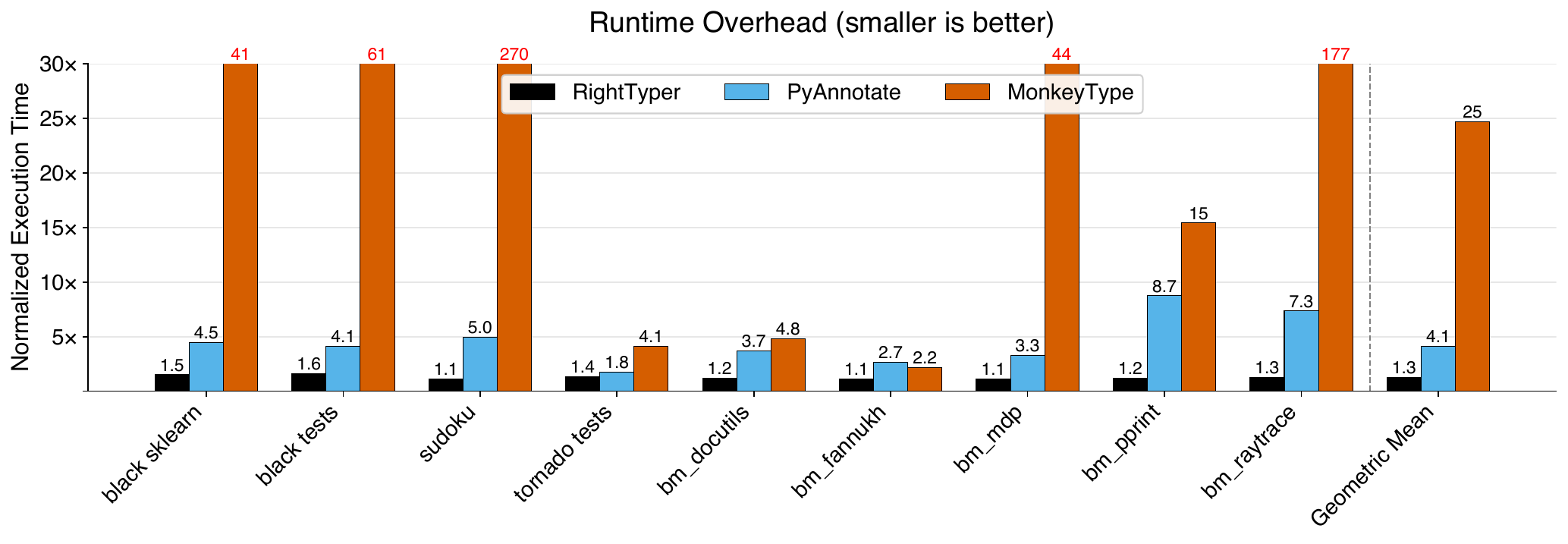}
    \caption{
        \textbf{[RQ2] \righttyper{} is much faster than existing dynamic tools}: across our benchmark suite, while PyAnnotate and MonkeyType incur high overheads (of \pn{benchmarks.overhead.PyAnnotate.geomean}{4.11}$\times$ and \pn{benchmarks.overhead.MonkeyType.geomean}{24.71}$\times$), \righttyper{}'s overhead remains about \pn{benchmarks.overhead.RightTyper.geomean_overhead_pct}{27}\%.
    }
    \label{fig:execution-time}
\end{figure*}

To evaluate \righttyper{}’s runtime performance, we measure the time it requires to collect type information while running a suite of benchmark applications, normalized against each application's baseline runtime.
We perform the same measurements for two existing dynamic type inference tools, MonkeyType and PyAnnotate; given the limited number of available dynamic inference tools, we include PyAnnotate.
We execute all tools with their default configurations.

We draw benchmarks from three sources.
To represent real-world Python applications, we run the \verb|black| code formatter on the large \verb|scikit-learn| package and execute a Sudoku solver~\cite{sudoku-solver}.
We also run the test suites of two widely used Python packages, \verb|black| and \verb|tornado|.
Finally, we include benchmarks drawn from Python’s benchmark suite~\cite{pyperformance}.

We run all experiments using Python 3.12.8 on a 10-core 3.7GHz Core~i9 system with 64GB of RAM and SSD storage, running Linux 6.5.6.
With the system otherwise idle, we run each experiment five times and report the median execution time.

\begin{table*}[t]
\centering
\caption{\textbf{[RQ2] \righttyper{} is fast:}
    While Type4Py and QuAC are faster, \righttyper{}'s execution time on the real-world code from \S\ref{eval-real-world} remains within reasonable bounds, matching pytype and faster than GPT-4o, in contrast to MonkeyType, which takes over 4.5 hours to complete.
    \label{table-real-world-timing}
}
\label{tab:tool-timing}
\begin{tabular}{l r r r r r r}
\toprule
Metric & Type4Py & QuAC & MonkeyType & pytype & GPT-4o & RightTyper \\
\midrule
TypeSim & 29.6\% & 28.0\% & 57.5\% & 19.8\% & 72.6\% & \textbf{77.2\%} \\
\midrule
Time (h:mm) & \textbf{0:05} & 0:06 & 4:44 & 0:14 & 0:22 & 0:08 \\
\bottomrule
\end{tabular}
\\[0.5ex]
{\small Larger TypeSim are better; shorter times are better.}
\end{table*}

Figure~\ref{fig:execution-time} shows the results.
Across all benchmarks, \righttyper{} consistently outperforms MonkeyType and PyAnnotate in terms of overhead.
\righttyper{} incurs a maximum slowdown of \pn{benchmarks.overhead.RightTyper.max}{1.6}$\times$ (geometric mean: \pn{benchmarks.overhead.RightTyper.geomean}{1.27}$\times$), compared to PyAnnotate's \pn{benchmarks.overhead.PyAnnotate.max}{8.7}$\times$ (geometric mean: \pn{benchmarks.overhead.PyAnnotate.geomean}{4.11}$\times$) and MonkeyType's \pn{benchmarks.overhead.MonkeyType.max}{269.6}$\times$ (geometric mean: \pn{benchmarks.overhead.MonkeyType.geomean}{24.71}$\times$).
While these results demonstrate that \righttyper{} is substantially faster than prior dynamic approaches, dynamic tools typically require execution under multiple scenarios to achieve sufficient code coverage.
To assess whether \righttyper{} is fast enough for practical use, we compare the running times of the evaluated tools on the real-world code described in \S\ref{eval-real-world}.

Table~\ref{table-real-world-timing} reports the previously achieved TypeSim scores alongside running times.
Type4Py and QuAC are the fastest, completing in about \pn{derived.type4py_time_minutes}{5} and \pn{derived.quac_time_minutes}{6} minutes respectively, followed by \righttyper{} at \pn{derived.rt_time_minutes}{8} minutes.
\texttt{pytype} and GPT-4o complete in \pn{derived.pytype_time_minutes}{14} and \pn{derived.gpt4o_time_minutes}{22} minutes, respectively; for GPT-4o, increased parallelism could likely reduce its execution time.
MonkeyType is by far the slowest, requiring \emph{nearly five hours}.
We conclude that \righttyper{} is fast enough for practical use.

\begin{conclusion}
\textbf{[RQ2] Summary:}
\righttyper{} incurs only $\sim$\pn{benchmarks.overhead.RightTyper.geomean_overhead_pct}{27}\% overhead---$\sim$\pn{derived.speedup_rt_vs_mt}{19}$\times$ faster than MonkeyType---and completes execution on real-world code in roughly eight minutes, competitive with static and AI-based approaches and fast enough for practical use.
\end{conclusion}

\subsection{[RQ3] Sampling Effectiveness}\label{eval-sampling}

\begin{table}[t]
\centering
\caption{\textbf{[RQ3] \righttyper{}'s trace sampling is effective:}
    MonkeyType's Bernoulli sampling degrades accuracy with modest speed gains, while \righttyper{}'s Poisson-timed sampling achieves both the highest accuracy and lowest execution time.
    \label{table-sampling}
}
\begin{tabular}{l r r r r}
\toprule
\multirow{2}{*}{Metric} & \multicolumn{3}{c}{MonkeyType} & \multirow{2}{*}{RightTyper} \\
& unsampled & $p{=}0.01$ & $p{=}0.001$ & \\
\midrule
TypeSim & 57.5\% & 37.4\% & 22.6\% & \textbf{77.2\%} \\
\midrule
Time (h:mm) & 4:44 & 3:05 & 3:04 & \textbf{0:08} \\
\midrule
Total traces & 126{,}377{,}033 & 1{,}463{,}432 & 151{,}559 & \textbf{218{,}797} \\
Unique functions\hspace{4em} & 3{,}869 & 2{,}318 & 1{,}391 & \textbf{3{,}895} \\
\bottomrule
\end{tabular}
\\[0.5ex]
{\small Larger TypeSim and more functions are better; shorter times and fewer traces are better.}
\end{table}

\parheading{Event Sampling}
A key design choice in \righttyper{} is its use of Poisson-timed event sampling to reduce overhead while preserving annotation quality (\S\ref{approach-instrumentation}).
To assess the importance of this design, we compare it against MonkeyType's Bernoulli sampling, which records all types from a random subset of calls.
We evaluate MonkeyType with sampling rates $p{=}0.01$ and $p{=}0.001$ on the same real-world code described in \S\ref{eval-real-world}.

Table~\ref{table-sampling} reports the results, including the number of \emph{total traces} (individual call observations recorded) and \emph{unique functions} (distinct functions for which at least one trace was recorded).
MonkeyType's TypeSim degrades substantially under sampling, dropping from \pn{sampling.monkeytype_unsampled.typesim}{57.5}\% without sampling to \pn{sampling.monkeytype_p100.typesim}{37.4}\% at $p{=}0.01$ and \pn{sampling.monkeytype_p1000.typesim}{22.6}\% at $p{=}0.001$.
Despite this degradation, execution time remains high---at roughly three hours for both sampling rates---suggesting that instrumentation and related overhead dominate MonkeyType's running time in these regimes.

In contrast, \righttyper{}’s Poisson-timed sampling achieves the highest TypeSim (\pn{sampling.righttyper.typesim}{77.2}\%) while also yielding the lowest execution time (\pn{derived.rt_time_minutes}{8} minutes).
It records only ${\sim}$\pn{derived.rt_traces_approx_k}{219}{,}000 traces---over \pn{derived.traces_ratio_mt_vs_rt}{600}$\times$ fewer than unsampled MonkeyType’s \pn{derived.mt_traces_approx_m}{126} million and \pn{derived.traces_ratio_p100_vs_rt}{6.7}$\times$ fewer than MonkeyType sampled at $p{=}0.01$.
Despite recording far fewer traces, \righttyper{} covers more functions: \pn{sampling.righttyper.functions}{3{,}895} versus \pn{sampling.monkeytype_unsampled.functions}{3{,}869} for unsampled MonkeyType and \pn{sampling.monkeytype_p100.functions}{2{,}318} at $p{=}0.01$.
Relative to unsampled MonkeyType, the improvement stems from \righttyper{}'s broader language support, including nested functions that MonkeyType does not handle.
Relative to sampled MonkeyType, which independently samples each function call with equal probability, \righttyper{} mitigates bias toward frequently executed (``busy'') functions, thereby improving functional diversity in the recorded traces.
Overall, these results demonstrate that \righttyper{}’s sampling strategy effectively balances overhead and accuracy.

\parheading{Container Sampling}
\righttyper{} tracks containers and either scans or samples their contents to infer full container types (\S\ref{container-sampling}).
To assess the effectiveness of this approach, we further instrument \righttyper{} to log every container observation---including the action taken (full scan, sampling, or spot-check), the number of samples drawn, and the types found---and to perform an exhaustive ground-truth scan of every container encountered.
We then re-execute \righttyper{} on the same real-world code from Section~\ref{eval-real-world} and compare the types that it finds against the ground truth.

\righttyper{} records \pn{derived.container_obs_millions}{6.6}M container observations, of which \pn{container_sampling.small_pct}{97.4}\% involve small containers ($\le 32$ entries) that are fully scanned.
The remaining \pn{derived.container_large_pct}{2.6}\% are large containers that trigger adaptive sampling.
Overall, \righttyper{} achieves perfect recall in \pn{container_sampling.perfect_recall_pct}{99.6}\% of observations and a mean recall of \pn{container_sampling.mean_recall_pct}{99.83}\%.
On large containers, where sampling replaces exhaustive scanning, it attains \pn{container_sampling.large_perfect_pct}{85.2}\% perfect recall with a mean recall of \pn{container_sampling.large_mean_recall_pct}{93.8}\%.

Sampling also yields substantial efficiency gains.
On average, \righttyper{} draws \pn{container_sampling.efficiency.samples_mean}{26} samples from large containers---mean size \pn{container_sampling.efficiency.size_mean}{837} entries---reducing total element inspections by \pn{container_sampling.savings_all_pct}{93.3}\% relative to exhaustive scanning---such as that performed by MonkeyType.
Following the initial observation, \righttyper{}'s change-detection mechanism---which triggers re-sampling when a container's size changes or when a spot check encounters a new type---detects \pn{container_sampling.detection_rate_pct}{74.3}\% of ground-truth type changes with a median latency of one observation.

\begin{conclusion}
\textbf{[RQ3] Summary:} \righttyper{}’s sampling strategy effectively balances overhead and accuracy. Poisson-timed event sampling covers \pn{derived.funcs_pct_more_rt_vs_p100}{68}\% more functions than MonkeyType’s Bernoulli sampling (at $p{=}0.01$) while recording \pn{derived.traces_ratio_p100_vs_rt}{6.7}$\times$ fewer traces.
Container sampling achieves perfect recall in \pn{container_sampling.perfect_recall_pct}{99.6}\% of observations while reducing element inspections by \pn{container_sampling.savings_all_pct}{93.3}\% relative to exhaustive scanning, demonstrating that adaptive sampling preserves accuracy while substantially lowering cost.
\end{conclusion}

\subsection{[RQ4] Good--Turing Estimation}\label{eval-good-turing}
\begin{figure}[t]
    \centering
    \includegraphics[width=\linewidth]{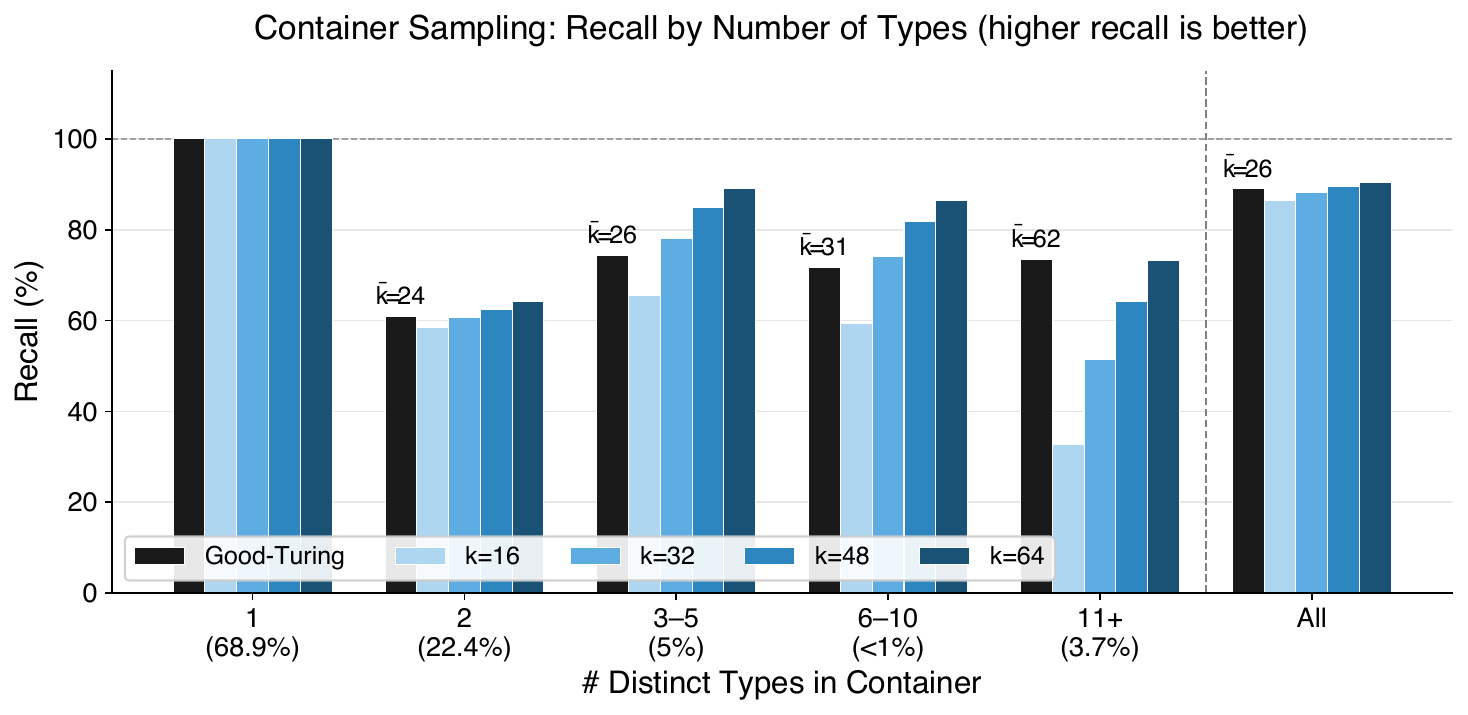}
    \caption{\textbf{[RQ4] Good--Turing adapts sampling effort to container diversity:}
        it matches random~$k{=}48$ in recall while drawing \pn{derived.gt_savings_pct}{47}\% fewer total samples, saving on single-type containers and investing more on diverse ones.
        Percentages indicate container observations in that bucket.
        Annotations above Good--Turing bars show the mean number of samples drawn ($\bar{k}$).
    \label{fig:good-turing}
    }
\end{figure}

To evaluate the Good--Turing stopping criterion used in container sampling (\S\ref{container-sampling}), we compare it against a baseline that draws a fixed number $k$ of random samples (with replacement) from each large container.
The additional instrumentation described in Section~\ref{eval-sampling} provides ground-truth type counts for all containers, enabling us to compute expected recall analytically.
The probability of observing type $t$ at least once in $k$ draws with replacement, and the expected recall (fraction of distinct types observed), are given by
$$P(\text{see type } t) = 1 - \left(1 - \frac{f_t}{N}\right)^{\!k}, \qquad E[\text{recall}] = \frac{1}{|T|}\sum_{t \in T} P(\text{see type } t)$$
where $f_t$ is the frequency of type $t$, $N$ is the container size, and $T$ denotes the set of distinct types.          

Figure~\ref{fig:good-turing} shows the results grouped by the number of ground-truth types per container.
Good--Turing's adaptive strategy concentrates effort where it matters most.
On homogeneous containers (1 type; \pn{good_turing.buckets.1.pct}{68.9}\% of cases), all strategies achieve perfect recall; Good–Turing draws the configured minimum of \pn{good_turing.buckets.1.gt_mean_k}{24}~samples.
On moderately polymorphic containers (2–10 types; \pn{derived.gt_moderate_pct}{27.4}\%), Good–Turing performs comparably to fixed $k{=}32$ (\pn{derived.gt_moderate_recall}{63.3}\% vs.\ \pn{derived.k32_moderate_recall}{63.8}\% in aggregate recall).
On highly polymorphic containers (11+ types; \pn{good_turing.buckets.11+.pct}{3.7}\%), Good–Turing draws a mean of \pn{good_turing.buckets.11+.gt_mean_k}{62}~samples and achieves \pn{good_turing.buckets.11+.gt_recall}{73.5}\% recall, comparable to $k{=}64$ (\pn{good_turing.buckets.11+.k64_recall}{73.2}\%).
The lower mean sample count in this bucket reflects the union-size cap (\S\ref{approach}): once the number of observed types exceeds the maximum union size, further sampling cannot improve the annotation (which collapses to \verb|Any|), so \righttyper{} stops early.
In aggregate, Good--Turing matches $k{=}48$ in recall (\pn{good_turing.aggregate.gt_recall}{89}\% vs.\ \pn{good_turing.aggregate.k48_recall}{89.5}\%) while drawing \pn{derived.gt_savings_pct}{47}\% fewer total samples (\pn{derived.gt_total_approx_m}{1.39} vs.\ \pn{derived.k48_total_approx_m}{2.62} million), by avoiding over-sampling the dominant homogeneous containers.

\begin{conclusion}
\textbf{[RQ4] Summary:} Good–Turing estimation is highly effective as a stopping criterion for container sampling.
In aggregate, it matches the recall of fixed $k{=}48$ sampling while drawing \pn{derived.gt_savings_pct}{47}\% fewer total samples by adaptively allocating effort based on container diversity.
\end{conclusion}

\subsection{Threats to Validity}\label{threats}

\parheading{Internal Validity}
Our evaluation depends on the correctness of our TypeSim reimplementation and the ground truth used.
We mitigate this by releasing our implementation and TypeEvalPy corrections for community review.
The choice of sampling parameters ($k{=}5$, $\lambda{=}2\,\mathrm{Hz}$, $\tau{=}0.05$ for event sampling; minimum 24 samples, budget cap 128 for container sampling) may affect results; we use defaults that balance overhead and coverage based on preliminary experiments.

\parheading{External Validity}
Our real-world evaluation covers five Python packages spanning development tools, libraries, and a web framework; they exercise diverse language features and contain annotations refined through years of production use.
Results may differ for other domains (e.g., ML code), though \righttyper{}'s language-agnostic design imposes no domain-specific limitations.
Additionally, when execution is driven by test suites, \righttyper{}’s coverage depends on test quality: codebases with limited test coverage may yield incomplete annotations, while tests that exercise unintended behavior may result in annotations that reflect that behavior.

\parheading{Construct Validity}
TypeSim measures structural similarity via shared attributes, which may not fully capture semantic equivalence---it penalizes protocol-based annotations and does not recognize type parameter equivalence.
Developer-written annotations serve as ground truth but may contain errors or stylistic choices rather than the most precise types.

\section{Related Work}
There has been substantial research on type inference for dynamically typed languages, particularly JavaScript, Ruby, and Python.
This section concentrates primarily on prior work targeting Python.

\partopic{Static type inference}\label{static-type-inference}
Static approaches analyze program structure without executing it, relying on techniques such as abstract interpretation, data-flow analysis, and type constraint resolution to infer types~\cite{maia2012static, pysonar, 10.1145/2661088.2661101, 10.1145/3551349.3561150, DBLP:conf/cav/HassanUE018, 10.1145/3689783, monat_et_al:LIPIcs.ECOOP.2020.17}.

Static type \emph{checkers} such as \verb|mypy|, \verb|pyright|, \verb|pyre|, and \verb|pytype| must also perform type inference to enable type checking, though their inference may be limited, primarily focused on how types from existing annotations propagate through the code or are affected by operations~\cite{mypy, pyright, pyre, pytype}.
Among these tools, only \verb|pytype| and \verb|pyre| support adding type annotations.

While typically sound by design, static approaches are constrained by dynamic language features such as runtime type changes, reflection, monkey patching, and dynamic code generation, and often support only a subset of their target languages~\cite{10.1145/1806596.1806598, hityper, 10.1007/978-3-642-03237-0_17, DBLP:conf/cav/HassanUE018}.
Dynamic features can also lead static systems to conservatively over-approximate, inferring overly permissive types~\cite{ruby-dynamic, typewriter}.
In contrast to static approaches, \righttyper{} produces precise annotations grounded in actual runtime behavior for all executed code paths.

\partopic{AI-based type inference}
These approaches estimate the likely types of program elements from features extracted from the source code, often including natural language elements. 
Xu et al. apply a probabilistic graphical model to predict variable types based on \emph{hints} such as attribute accesses and variable names~\cite{10.1145/2950290.2950343}.
TypeWriter applies a neural network to both code and natural language information and verifies the predictions using a type checker~\cite{typewriter}.
Typilus and Type4Py leverage deep similarity learning to avoid using a fixed type vocabulary~\cite{typilus, type4py}.
PYInfer uses a neural network to infer variable types from the surrounding source code context~\cite{pyinfer}.
HiTyper combines static and machine learning inference by only applying the machine learning model where its type constraints resolution gets stuck~\cite{hityper}.
Abdelaziz et al. combine mining docstrings and approximate duck subtyping to infer function return types~\cite{abdelaziz-et-al-2022, duck-typing}.
TypeT5 fine-tunes CodeT5 for type inference~\cite{typet5, codet5}.
DLInfer extracts program slices for variables and applies a sequence model to predict function argument types~\cite{10.1109/ICSE48619.2023.00170}.
TypeGen prompts ChatGPT with chain-of-thought instructions to infer types~\cite{typegen, chain-of-thought}.
TIGER ranks generative model typing candidates with a similarity model~\cite{tiger}.
DeMinify, primarily targeting minified code recovery, uses dual-task learning between variable name and type prediction~\cite{deminify}.
RunTyper combines dynamic analysis with deep neural networks to validate and refine type predictions; it employs a modified Python interpreter to gather runtime type information~\cite{10.1145/3771544}.

While better able to cope with Python's dynamic nature than purely static approaches, AI-based methods are inherently unsound: their predicted types may not reflect actual program behavior.
By combining AI-based and static or dynamic inference, approaches such as TypeWriter, HiTyper, Typilus, and RunTyper can reject incorrect types---but risk leaving more of the code unannotated~\cite{typewriter, hityper}.
Furthermore, these approaches often only support a limited, small type vocabulary and perform poorly on rare types, making them less effective when encountering custom or evolving types~\cite{type4py}.
Unlike AI-based approaches, \righttyper{} produces annotations grounded in actual runtime behavior.

\partopic{Dynamic type inference}
Dynamic approaches track types encountered during program execution.
Tools such as MonkeyType (Instagram), PyAnnotate (Dropbox, developed by Python author Guido van Rossum), and \righttyper{} employ this strategy~\cite{monkeytype, pyannotate}.
A major limitation of prior dynamic approaches is the runtime overhead they impose.
Although both MonkeyType and PyAnnotate support sampling of function calls to mitigate overhead, their designs require some instrumentation to remain continuously active, incurring nontrivial cost.
MonkeyType implements Bernoulli random sampling (disabled by default), but even when enabled, this requires instrumentation on every call to decide whether to sample, and MonkeyType logs all collected data to a SQLite database.
Moreover, for each sampled call, MonkeyType recursively and exhaustively scans every element of every container argument---resulting in severe performance degradation for large containers and, in some cases, causing execution to abort due to timeouts or resource exhaustion.
In contrast, \righttyper{} uses a Poisson process: instrumentation is disabled except during brief capture windows, so most calls incur zero overhead.
Because capture windows occur at random times, every call has an equal probability of being observed, ensuring representative sampling without systematic bias.
PyAnnotate's deterministic strategy also avoids per-call decisions but can introduce substantial bias; \righttyper{} uses a Good--Turing estimator for containers, which is both efficient and statistically principled.

Both MonkeyType and PyAnnotate rely on Python's runtime type attributes, which are not always reliable.
MonkeyType prunes invalid type names, but this can leave code unannotated; PyAnnotate performs no validation and may emit annotations causing runtime errors.
\righttyper{} instead maps type objects to valid, public names.

Additionally, MonkeyType and PyAnnotate lack support for important language features: inferring variable types, handling keyword and variadic arguments, typing most non-container generics, and correctly typing inherited methods.
When types vary across executions, they construct simple unions that can be incorrect or overly permissive.
In contrast, \righttyper{} supports these features, recognizes recurring type patterns, and applies a suite of context-aware simplification strategies where appropriate (\S\ref{object-typing}--\ref{simplifying-types}).
As far as we are aware, no typing tool other than \righttyper{} supports annotating array dimensions (\S\ref{typing-generics}).

\section{Conclusion}
\Thispaper{} introduces \righttyper{}, a novel hybrid type inference approach for Python that produces accurate and precise type annotations.
By combining dynamic analysis with static analysis, type aggregation, and name resolution, \righttyper{} generates annotations of substantially higher quality than those produced by prior approaches.
Its principled, statistically guided adaptive sampling enables it to observe representative program behavior while maintaining low runtime overhead.
Together, these characteristics make \righttyper{} a practical and effective tool for annotating real-world Python programs, enabling developers to benefit from static type checking with minimal manual effort.

Future work includes extending \righttyper{} to additional language features and improving support for already annotated codebases.

\section*{Data Availability}

\ifanonymous
A replication package, including \righttyper{} sources, is available anonymized at \replicationurl{}.
We intend to make data publicly available upon acceptance.

\smallskip\noindent
\textbf{Note that the project name is unchanged from the original; conducting a web search is likely to compromise anonymity.}
\else
\righttyper{} is available as source code at \righttyperurl{}.
\fi


\clearpage
\bibliographystyle{ACM-Reference-Format}
\bibliography{references,emery}{}

\end{document}